# Hybrid Machine Learning Model of Extreme Learning Machine Radial basis function for Breast Cancer Detection and Diagnosis; a Multilayer Fuzzy Expert System

Sanaz Mojrian
Department of Information
Technology Mazandaran
Uni. of Science and Technology
Babol, Iran
0000-0003-3356-2942

Gergo Pinter
John von Neumann Faculty of
Informatics Obuda University
Budapest, Hungary
0000-0003-4731-3816

Javad Hassannataj Joloudari
Electrical and Computer
Engineering Faculty, University
of Birjand, Birjand, Iran
0000-0001-9374-2326

Imre Felde
Kalman Kando Faculty of
Electrical Engineering
Obuda University
Budapest, Hungary
0000-0003-4126-2480

Narjes Nabipour
Institute of Research and
Development, Duy Tan
University, Da Nang 550000,
Vietnam
0000-0003-3882-3179

Laszlo Nadai
Kalman Kando Faculty of
Electrical Engineering
Budapest, Hungary
0000-0001-8216-759X

Amir Mosavi
Kalman Kando Faculty of
Electrical Engineering
Budapest, Hungary
0000-0003-4842-0613

*Abstract*— **Mammography is often used as the most common laboratory method for the detection of breast cancer, yet associated with the high cost and many side effects. Machine learning prediction as an alternative method has shown promising results. This paper presents a method based on a multilayer fuzzy expert system for the detection of breast cancer using an extreme learning machine (ELM) classification model integrated with radial basis function (RBF) kernel called ELM-RBF, considering the Wisconsin dataset. The performance of the proposed model is further compared with a linear-SVM model. The proposed model outperforms the linear-SVM model with RMSE, R$^2$, MAPE equal to 0.1719, 0.9374 and 0.0539, respectively. Furthermore, both models are studied in terms of criteria of accuracy, precision, sensitivity, specificity, validation, true positive rate (TPR), and false-negative rate (FNR). The ELM-RBF model for these criteria presents better performance compared to the SVM model.**

*Keywords—hybrid machine learning, extreme learning machine (ELM), radial basis function (RBF), breast cancer, support vector machine (SVM),*

## I. INTRODUCTION

Breast cancer is among the most common disease of young women over the world [1-3]. Approximately 29.9% of mortality from cancer in women is due to breast cancer. The incidence of this disease is lower in developing countries than in developed countries, about 10% of women with breast cancer in Western countries. However, millions of people over the world are suffering from this life-threatening disease mortality [3,4]. But recently, mortality in developed countries has increased, with global data showing about one million and six hundred thousand new cases of breast cancer were reported in 2012. Seven hundred ninety-four thousand people in developed countries compared to 883,000 people in developing countries, though 198,000 mortalities in developed countries compared to 324,000 people in less developed countries [5]. The increasing global prevalence of the disease and the resulting death rate can be attributed to the significant and growing threat in the developing world. Breast cancer caused by growing out the rule is abnormal cells in the breast. Cancer can be a common phenomenon in developing countries, that is, unfortunately, due to changes in lifestyle as well as hormonal therapies. Early detection of the disease has a crucial role in reducing mortality. However, the early treatment also requires early detection of breast cancer at an initial stage. Most methods for detection of this disease are performed through surgery or mammography screening and are costly; countries with good health systems use screening techniques that will result in high financial costs [6].

Several studies use data mining and artificial intelligence techniques for diagnosis and detection [7-13]. In 1995, data mining and machine learning were used in decision-making to detect breast cancer [14]. Data mining and machine learning techniques had been in use to enhance the speed and accuracy of diagnosis [3,15]. However, increasing accuracy to reduce the RMSE error rate is considered to be a challenge, and this subject to research is being discussed. Therefore, in this study, we introduce the multilayer model and investigate the performance of two important and well-known models. We first label the dataset using fuzzy logic and then use the proposed ELM and SVM-based models to classify the best performance between the two models to detect benign and malignant breast cancer with the lowest error rate. To evaluate the performance of the proposed models, we used the confusion matrix. We also applied the 10-Fold cross-validation technique to obtain the average accuracy of the models. Once again, the dataset is divided into three groups of training, testing, and validation that is investigated evaluation criteria including accuracy, precision, sensitivity, specificity, validation, false-positive rate and false-negative rate for the ELM and SVM models that they are very effective to determine if the performance of these two models.

## II. LITERATURE REVIEW

The use of data science and data mining in technology and machine learning has significantly contributed to medical science. In recent years, many studies have been done on early detection and prevention of disease and the selection of suitable treatment for diseases [16]. Breast cancer of this rule is no exception, so that much study and research have been done and continues to be completed in this field. In this section, we briefly review several studies in the field of breast cancer that is done. One of the first works to do in this regard can be the decision tree C4.5 algorithm with a 10-fold cross-validation technique on the Wisconsin dataset to detect the benign tumors that the accuracy of detection for this algorithm was computed 94.74% [17]. The decision tree algorithm was also used to detect the disease, which was separated from benign and malignant masses of 92.97% accuracy [18]. Weighted NB is a new method for classifying breast cancer, which is considered the several performances of evaluation techniques, including sensitivity, specificity, and accuracy in [19]. According to the results of the experiments, the sensitivity, specificity and accuracy values were computed 99.11%, 98.25%, and 98.54%, respectively. Useful information extraction and tumor detection are developed using K-means and support vector machine (K-SVM) algorithms in combination. The detection accuracy of the proposed method was obtained at 97.38% [20]. In another study, a new knowledge-based system was used to classify breast cancer using clustering techniques called Expectation-Maximization (EM) and classification and regression trees (CART) to generate fuzzy rules on the Wisconsin dataset. The accuracy computed in this system based on the fuzzy reasoning method was estimated at 93.20% [21]. Other methods for detection of breast cancer include the supervised learning algorithm using a fuzzy clustering algorithm 97.57% accuracy [22]. Authors in [23] utilized a variety of classification methods, including SVM, probabilistic neural network (PNN), recurrent neural network (RNN), combined neural network (CNN), and Multilayer Perceptron Neural Network (MLPNN) for the detection of breast cancer, which they were achieved 99.45%, 98.61%, 98.15%, 97.40%, and 91.92%, respectively of accuracy. The authors were also able to identify benign and malignant tumors with 98.53% accuracy using the least square support vector machine (LS-SVM) method [24]. Recently, three data mining algorithms including Naive Bayes, RBF network, J48, have been used to predict breast cancer. For this purpose, a dataset with 683 samples from three continents, including Asia, Africa, and Latin America was exploited. 10- fold Cross-Validation technique was also used to evaluate the models. The results denote that the Naïve Bayes model is the best predictor compared to the RBF network and j48 models with an accuracy of 97.36%, 96.77%, and 93.94%, respectively [25].

## III. MATERIALS AND METHODS

### A. Data

The study used the Wisconsin Breast Cancer Dataset (WBCD) for women in the UCI machine learning dataset [26]. This dataset contains 699 records with 16 missing values for the Bare Nuclei feature taken by Dr. W.H. Wolberg from aspiration needle at the University of Wisconsin Madison Hospital of human breast cancer tissue. Depending on the values of these features, the benign and malignant mass is determined. Also, the WBCD contains nine features in this dataset. Depending on the values of these features, the type of benign and malignant mass is distinguished. In the dataset with 16 missing values, we discard these missing values in our experiment, and we consider only the remaining 683 records. From the cleaned dataset, 444 records belonged to the benign class, and the remaining 239 records belonged to the malignant class. Two classes for benign and malignant breast cancer are classified as classes 2 and 4, respectively.

### B. Methodology

In this research, the proposed system consists of 6 modules, each module consisting of several phases. It's a multilayer system. Fig. 1, presents the methodology. In the following, we will fully describe each module and components performance in the respective phase. The first module investigates the dataset. This module is an initial dataset investigate module that is part of the pre-processing data module. The first module consists of 3 phases. The first phase: captures the data from the dataset. The second phase analyzes each detail of the data to enhance the quality of the data. The third phase prepares the data by deleting noise or non-content data to transfer to the next module. The flowchart of the 3 phases related to the first module is illustrated in Fig. 2. After an initial investigation, the confirmed dataset is stored in a log file, which is cleaned dataset from breast cancer. The purpose of creating the first module is to perform the suitable activities based on the steps mentioned on the dataset, and the result of these activities or processing will be transferred to the next module after being stored in a log file as input.

In the fuzzy breast cancer detector (FBCD) module, the fuzzy labeling is used to detect the breast cancer to label data. In other words, the second module is responsible for labeling the file log, namely fuzzy labeling. According to the second module, the purpose of the fuzzy system in analyzing a breast cancer dataset is to distinguish the benign and malignant outcome of the mass. Before starting the fuzzy discussion in this study, the input features must be fuzzification. A range of 1 to 10 is assigned to each feature that should be medically within the range. The input variables have been imported to the fuzzification phase. In this module, input variables are transformed into fuzzy linguistic variables to facilitate the modeling and extraction of medically acceptable results. Table. I illustrates a representation of the linguistic variables for breast cancer disease features with a range of 1 to 10 designed in this study. According to the dataset, as well as the fuzzy linguistic variables, we describe the fuzzy table of the nine features. Therefore, the procedure of the second module is accomplished through the fuzzy threshold system after the initial evaluation of the breast cancer dataset, input selection, data labeling, and detection of benign and malignant of a cancer mass.

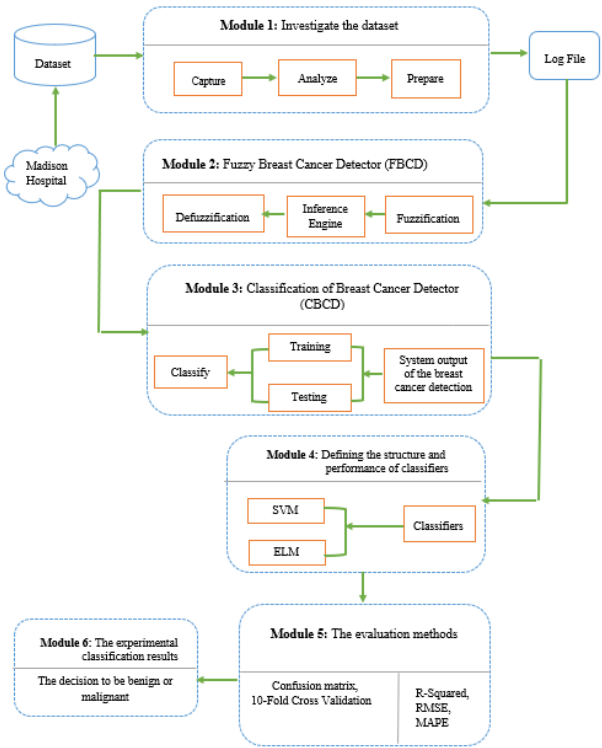

Fig. 1. Schematic representation of the research method

TABLE I.   LINGUISTIC VARIABLES FOR BREAST CANCER DISEASE FEATURES WITH THE RANGE.

| Features | fuzzy linguistic variables | Range |
|---|---|---|
| Clump Thickness | (Low), (Medium), (High) | 1-10 |
| Uniformity of Cell Size | (Low), (High) | 1-10 |
| Uniformity of Cell Shape | (Low), (High) | 1-10 |
| Marginal Adhesion | (Low), (High) | 1-10 |
| Single Epithelial Cell Size | (Low), (High) | 1-10 |
| Bare Nuclei | (Low), (High) | 1-10 |
| Bland Chromatin | (Low), (High) | 1-10 |
| Normal Nucleoli | (Low), (Medium), (High) | 1-10 |
| Mitoses | (Low), (High) | 1-10 |

The first phase is feature extraction for fuzzification to extract the important features from the Wisconsin dataset. Several important features in the detection of breast cancer are analyzed. The dataset is included nine features for breast cancer such as Clump Thickness, Uniformity of Cell Size, Uniformity of Cell Shape, Marginal Adhesion, Single Epithelial Cell Size, Bare Nuclei, Bland Chromatin, Normal Nucleoli, and Mitoses. In the proposed system, all of these nine features are considered as crucial input variables. The second phase is Fuzzification where every crisp input value must be transformed into a fuzzy variable with respect to the linguistic variables. The fuzzy set functions are assigned values between 1 and 10, indicating the membership function of an element in a specific set. The third phase is the fuzzy interface engine and rule base, where the database stores fuzzy rules and knowledge of the rules used in the fuzzy inference engine to obtain a new reality. The fourth phase is defuzzification.

According to Table. I, the fuzzy linguistic variables include "Low", "Medium" and "High". The numbers range is considered from 1 to 10 medically. It should be noted that we use MATLAB software and a fuzzy toolbox to implement the methods. The fuzzification process for the proposed detection system is shown in Fig. 3.

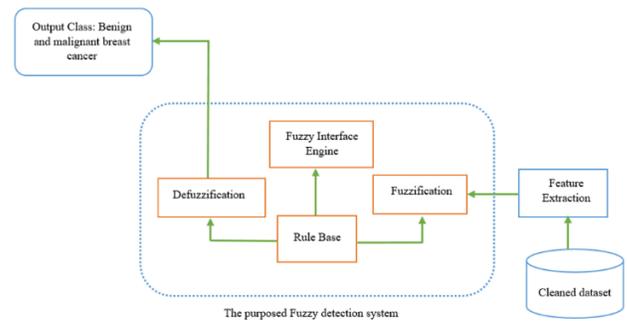

Fig. 3. Architecture for fuzzification of the proposed fuzzy detection System of breast cancer.

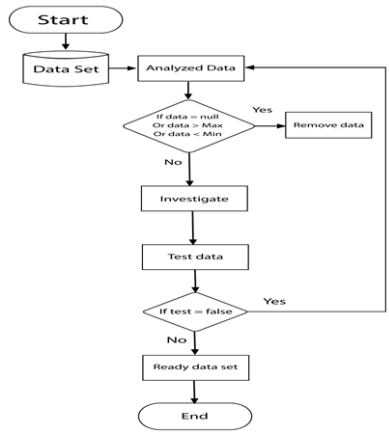

Fig. 2. Flowchart for the first module

In the fuzzification phase, the input variables, which are numeric and absolute, are transformed into linguistic variables according to the fuzzy inference engine and fuzzy rules to be understood, finally being numerically labeled with defuzzification [27,28]. As shown in Fig. 3, the second module consists of several phases.

Due to the fuzzy rules applied to the inference engine and the linguistic variables in output, the defuzzification phase is to transform the linguistic variables into crisp values. It actually does numeric labeling. The fifth phase concerns with the output class where the intelligent analysis and decision-making phase by the proposed fuzzy system as it deals with the labeling phase, defuzzification output and with regard to the 9 crucial features of breast cancer, as well as the fuzzy rules, the relative inputs are analyzed by the proposed fuzzy system, so that each data has been its suitable label. The labeled data is then ready to be transferred to the third module after analyzing the fuzzy system. The purpose of the third module is to use classification algorithms to detect and separate benign and malignant breast cancer tumors. Fig. 4 shows the implementation phases for the proposed fuzzy system.

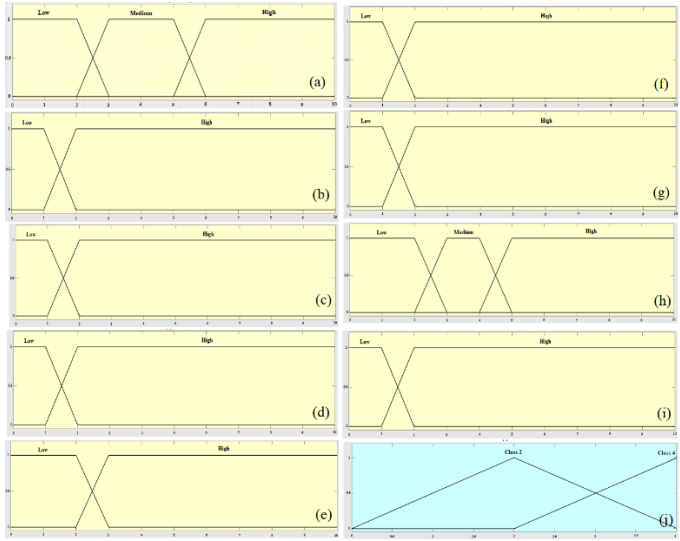

Fig. 4. Membership function plots for input variables: (a) Clump Thickness, (b) uniformity of cell size, (c) uniformity of cell shape, (d) marginal adhesion, (e) single epithelial cell size, (f) bare nuclei, (g) bland chromatin, (h) normal nucleoli, (i) mitoses and (j) the output variable (fuzzy set for selector field as class 2 and 4).

The third Module is the Classification of Breast Cancer Detector (CBCD). The third module includes the phase of classifying the breast cancer dataset [29] by using an SVM and a neural network model based on extreme learning capability namely Extreme Learning Machine (ELM) to detect benign or malignant type of cancer mass that in this module, dataset is divided into three groups: training, testing, and validation. The first group is training. The system training process can be considered as a search to determine the local minimum of a cost function (typically the sum of squares of error or the mean of squares of error). In the training phase, using the training dataset, the related model is constructed [30-31]. In this study, after preparing the new data structure file, 70% of the data is applied to SVM and neural network-based ELM classifiers for training. To implement the proposed model in MATLAB software, an input matrix, including nine features, is inserted into the selected model. The target matrix represented two classes of benign and malignant mass with 2 and 4 labels, respectively. If the type of cancer belongs to the relevant class, its rows are filled with 2 and 4 values. The second group is testing. At this phase, after the trained models, the variables that are considered for testing the models are applied to the proposed system and their outputs for computing are investigated and analyzed. The testing dataset is used to compute the accuracy of the constructed model [30-31]. In this study, 20% of the dataset is used to test and to evaluate the results of the models. The third group is the validation. In this study, we used the validation field to compute the validation of the two SVM and ELM classification models. For this purpose, we consider 10% of the data for the validation of models. The fourth module is defining the structure and performance of the classifiers. We define the structure and performance of each of the SVM and ELM classifiers in this module. SVM is a supervised learning algorithm based on statistical learning theory and structural risk minimization that it is introduced by Vapnik [32]. The SVM is generally used for problems where there are two classes so that SVM uses an optimized linear separator hyperplane to create classes, but not all problems in the main input space can be linearly separated. Therefore, SVM is a classifier that maps the linear training patterns into nonlinear patterns in high dimensional feature space. The linear decision-making function for SVM is as follows;

$$f(x) = sign(w.\emptyset(x) + b) \qquad (1)$$

The distance between the two classes called margin, in the new feature space is $\frac{2}{||w||}$. The SVM model attempts to maximize the margin and minimize the error of the training data. In this model, the optimization problem is as follows.

$$Minimize\ L_{Primal-SVM} = \frac{1}{2}||w||^2 + C\sum_{i=1}^{N}\xi_i \qquad (2)$$

The user-defined "C" parameter turn-out a balance between margin size and training error. The $x_i$ data that is $t_i(w.\phi(x_i) + b) = 1$ is called the support vector. Using the KKT theorem, the dual optimization problem of the previous problem is computed as follows [33].

$$Minimize\ L_{Dual-SVM} = \frac{1}{2}\sum_{i=1}^{N}\sum_{j=1}^{N}t_it_j\alpha_i\alpha_jK(x_i,x_j) - \sum_{i=1}^{N}\alpha_i \qquad (3)$$

$$subject\ to: \sum_{i=1}^{N}t_i\alpha_i = 0\ ,\ 0 \le \alpha_i \le C, i = 1,\dots,N$$

ELM first proposed by Huang et al. [34]. ELM operates on generalized Single hidden layer feedforward networks (SLFN). In the ELM model, the hidden layer does not need to be adjusted, and the functions of this layer, which is a feature transfer to the new space, are already specified. SVM models, polynomial neural networks, Radial Basis Function (RBF), and single-layer feedforward models are a special state of this model. In kernel-based SVM, PSVM, and LS-SVM methods, it is first assumed that data is transferred to kernel space. This space "$\emptyset(x)$" that maps data from the current space to the new space is usually unknown and uses the equivalent kernel. This is performed with the purpose of generalization of these models to the learning of nonlinear areas. In the current data space, these models are only able to find linear areas. Transferring to the kernel space, we actually go into a space that contains nonlinear elements of the previous space. The decision function that is ultimately learned in these three methods mentioned above is as follows;

$$f(x) = sign(\sum_{i=1}^{N}\alpha_it_iK(x,x_i) + b) \qquad (4)$$

Where "$x_i$" is the ith training data and "$t_i$" is the label of this training data (assuming $t_i \in \{-1,+1\}$ ). The function "$K(x,x_i)$" of the kernel is between the new data "$x$" and the training data "$x_i$". There is also the bias "b" for the decision-making function. The ELM model was first created for SLFNs and it was later developed to generalized SLFNs. The decision function in this model is as follows;

$$f(x) = h(x)\beta = \sum_{i=1}^{L}\beta_i h_i(x) \quad (5)$$

Where $h(x) = [h_1(x), ..., h_L(x)]$ is a transformation to a new feature space performed by the hidden layer. The vector $\beta\beta = [\beta_1, ..., \beta_L]$ is the weight vector between the latent layer and the output layer, which must be learned. Here we can model the SVM as an ELM that is as follows;

$$h_i(x) = t_i K(x, x_i), i = 1,2, ...., N \quad, h_0(x) = 1 \quad (6)$$

Therefore, by considering the above transformation and the optimization problem for the ELM model such that the weights equal to $\beta = [b, \alpha_1, ..., \alpha_N]$ are found, we can find the SVM model equivalent to SVM. As with SVM-based models, if the feature mapping is unknown to us, kernels can also be used in ELM models. In this state, the ELM kernel is defined as follows.

$$\Omega_{ELM} = HH^T : \Omega_{ELM\ i,j} = h(x_i).h(x_j) = K(x_i, x_j) \quad (7)$$

In this state, the model output function can be computed by the kernel as follows [26].

$$f(x) = h(x)H^T\left(\frac{I}{C} + HH^T\right)^{-1} T = \begin{bmatrix} K(x, x_1) \\ \vdots \\ K(x, x_N) \end{bmatrix}^T \left(\frac{I}{C} + \Omega_{ELM}\right)^{-1} T \quad (8)$$

Also, in this study, the radial basis function (RBF) is selected as the kernel function as denoted below, where σ is the radius of the base function of the kernel.

$$K(xi, xj) = e^{\frac{-\|x_i - x_j\|}{2\sigma^2}} \quad (9)$$

In this study, the evaluation metrics of MAPE, RMSE, $R^2$, confusion matrix and 10-fold cross-validation methods are used. $R^2$ is the quantity called the coefficient of determination so that it is computed as the ratio of the changes defined to the total changes. In other words, this coefficient indicates that "how much of the dependent variable changes are affected by the relevant independent variable" [35]. The coefficient of determination, whatever be closer to "1", the relative model is more efficient. The value of "$R^2$" is computed as follows.

$$R2 = \frac{\Sigma\{(x_i - \bar{x}_i) * (x_j - \bar{x}_j)\}}{\sqrt{\Sigma(x_i - \bar{x}_i)^2 * \Sigma(x_j - \bar{x}_j)^2}} \quad (10)$$

In (10), $x_i$, $\bar{x}_i$, $x_j$ and $\bar{x}_j$ are denoted the values estimated by the model, the average of the values estimated by the model, the actual values and the average of the actual values, respectively. $R^2$ represents the correlation between the experimental and the estimated data, i.e. the value of the coefficient of determination of closely "1" denotes better model efficiency and the higher correlation between the experimental and the estimated data.

The root-mean-square error (RMSE) is a useful error criterion for accuracy measurement [36]. The difference is between the value predicted by the model and the actual value. Unlike R2, the RMSE should be closer to "0".

$$\text{RMSE} = \sqrt{\frac{\Sigma(x_j - x_i)^2}{n}} \quad (11)$$

where $n$ is the number of records. Also, $x_j$ is the actual values and $x_i$ is the values estimated by the model.

Mean Absolute Percentage Error (MAPE), also called mean absolute percentage deviation (MAPD), is a criterion for measurement of predictive accuracy as a percentage in statistics and it is defined by the following equation [37].

$$\text{MAPE} = \frac{100\%}{n}\sum_{t=1}^{n}\left|\frac{A_t - F_t}{A_t}\right| \quad (12)$$

According to (12), $A_t$ is the actual value and $F_t$ is the predicted value.

To evaluate the performance of the SVM and ELM models in more detail, the confusion matrix is used as illustrated in TABLE. II. Using this matrix, the value of each criterion is computed and then the results are compared. This matrix is a useful tool for analyzing the performance of classification models in detecting data in different classes [15,38].

TABLE II. CONFUSION MATRIX IN THIS STUDY.

| The predicted class | Tumor (malignant) | Tumor (benign) |
|---|---|---|
| Malignant | True Positive | False Positive |
| Benign | False Negative | True Negative |

For the evaluation of the expressed models, a confusion matrix is used which consists of 4 elements [15], i.e., TP: True positive indicates the number of correctly classified positive predictive records. FP: False positive, indicating the number of positive predictive records that are incorrectly classified as negative. FN: False-negative, indicating the number of negative predictive records that are incorrectly classified as positive.
TN: True negative indicates the number of correctly classified negative predictive records. The 10-Fold Cross-Validation [39], divided data into ten subsets. Then they are processed ten times. In this process, seven datasets for training, 2 subsets for testing and one remaining subset for validation are used. Therefore, the average accuracy computed with ten iterations is computed using 10-fold cross-validation.

IV. RESULTS

The results are extracted to decide regarding be benign or malignant of cancer mass in this study. A comparison of the results of the classification models according to RMSE, $R^2$, and MAPE evaluation criteria is performed for three phases including training, testing, and validation that these results are denoted in Tables III -V. Further, a comparison of results using the confusion matrix is given.

TABLE III. EVALUATION CRITERIA FOR THE TRAINING PHASE.

| RMSE | $R^2$ | MAPE | Model |
|---|---|---|---|
| 0/4596 | 0/7956 | 0/0978 | **Linear-SVM** |
| 0/3224 | 0/9147 | 0/0764 | **ELM-RBF** |

TABLE IV. EVALUATION CRITERIA FOR THE TESTING PHASE.

| RMSE | $R^2$ | MAPE | Model |
|---|---|---|---|
| 0/2649 | 0/8761 | 0/0693 | **Linear-SVM** |
| 0/1719 | 0/9374 | 0/0539 | **ELM-RBF** |

TABLE V. EVALUATION CRITERIA FOR VALIDATION PHASE.

| RMSE | $R^2$ | MAPE | Model |
|---|---|---|---|
| 0/2157 | 0/7929 | 0/0582 | **Linear-SVM** |
| 0/1584 | 0/8643 | 0/0417 | **ELM-RBF** |

After observing the results of the evaluation criteria, we concluded that ELM-RBF had the best classification result. As stated earlier the has been divided to 70% for training, 20% for testing, and 10% for validation of dataset. Furthermore, several criteria were used to compare the models, including accuracy in training and testing, precision, sensitivity, specificity, validation, false-positive rate, and false-negative rate. The (13)-(18) are calculated as follows.

Specificity = True Negative Rate (TNR) = TN / TN + FP   (13)
Sensitivity = True Positive Rate (TPR) = TP/TP + FN   (14)
Accuracy = TP + TN / TP + TN + FP + FN   (15)
Precision = TP / TP+FP   (16)
F-measure = 2*Precision*Recall / Precision + Recall   (17)
FPR = 1- Specificity   (18)
FNR = 1- Sensitivity   (19)

The performance results of two classification algorithms were compared on 9 features including Clump Thickness, Uniformity of Cell Size, Uniformity of Cell Shape, Marginal Adhesion, Single Epithelial Cell Size, Bare Nuclei, Bland Chromatin, Normal Nucleoli and Mitoses. We found acceptable results for breast cancer in the Wisconsin dataset. The results of the experiment using the confusion matrix for criteria of accuracy, precision, sensitivity, specificity, validation, false positive rate and false negative rate demonstrate that the proposed ELM model has better performance than SVM model. We used the 10-fold cross-validation technique in this study so that the average accuracy computed for the models using this evaluation technique demonstrates that the proposed model has a detection accuracy of 98.05. While the accuracy of the SVM model is 90.56%. A comparison of the results for the two models for the criteria mentioned in Fig. 5 is shown.

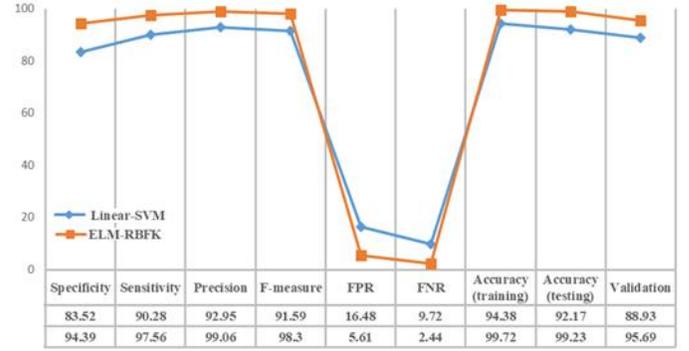

Fig. 5. Comparison of classification results according to the evaluation criteria of models.

According to the analysis demonstrated in Fig. 5, we found that the ELM model has better performance than the SVM model with 9 features in all criteria, so that for the ELM-RBF model the accuracy is computed 99.23% in the test phase and the accuracy of this model is computed 99.72% in the training phase. However, for the SVM model, the accuracy was computed 92.17% and 94.38% in the test phase and in the training phase, respectively. Also, the validation measurement of the ELM and SVM models were computed 95.69 and 88.93, respectively. Besides, the FPR measurement for the ELM-RBF model is 5.61 versus the SVM model which is 16.48 and the FNR measurement for the ELM and SVM models is 2.44 and 9.72, respectively.

V. CONCLUSION

A Multilayer Fuzzy Expert System of ELM-RBF proposed on the Wisconsin dataset for breath cancer detection. This model was compared with the SVM model. The performance of the classifiers was evaluated according to fuzzy labeled data. Then, the best classifier was distinguished to be benign and malignant of cancer mass. With this hybrid method, an efficient prediction model was determined to identify the type of cancer mass. For this purpose, the cleaned dataset consists of 683 records divided into 3 sections; training, testing and validation so that 70% of the data for training, 20% for testing and 10% for model validation were considered. We used three criteria for evaluation such as RMSE, $R^2$ and MAPE to the detection of breast cancer. The results demonstrate that the proposed model has the best performance than SVM according to tables 2 to 4. Also, the average accuracy computed using the 10-fold cross validation technique for the ELM-RBF model has been 98.05. However, the accurate measurement for the SVM model has been 90.56. Another important achievement of this study is that we use the confusion matrix to evaluate the relevant models, which demonstrates that the proposed ELM-RBF model is better than the SVM model in terms of criteria of accuracy, precision, sensitivity, specificity, validation, false positive rate and false negative rate. So, by dividing the dataset into three groups of training, testing and validation, the accuracy of training and testing as well as validation for the ELM-RBF model were 99.72, 99.23 and 95.69%, respectively, and these measurements for the SVM model were 94.38 for training, 92.17% for testing and 88.93% for validation. Among these criteria, the FPR criterion is more important than the FNR criterion for clinical centers so that the FPR measurement of 5.61 versus the SVM model of 16.48. The comparison of the mentioned classification results according to the evaluation criteria of models is provided. Therefore, the proposed fuzzy ELM-RBF model can replace medical invasive diagnostic methods, as it can distinguish patients who do not require sampling and testing with high accuracy. This method of early and accurate detection very crucial because it avoids the necessary complications and high costs in detecting the disease of breast cancer.

ACKNOWLEDGMENT

We acknowledge the financial support of this work by the Hungarian State and the European Union under the EFOP-3.6.1-16-2016-00010 project